# Observation of converse flexoelectric effect in topological semimetals

Hidefumi Takahashi[1,2,*], Yusuke Kurosaka[1], Kenta Kimura[1], Akitoshi Nakano[3] and Shintaro Ishiwata[1,2,†]

[1]*Division of Materials Physics and Center for Spintronics Research Network (CSRN), Graduate School of Engineering Science, Osaka University, Osaka 560-8531, Japan*
[2]*Spintronics Research Network Division, Institute for Open and Transdisciplinary Research Initiatives, Osaka University, Yamadaoka 2-1, Suita, Osaka, 565-0871, Japan*
[3]*Department of Physics, Nagoya University, Nagoya 464-8602, Japan*

*takahashi.hidefumi.es@osaka-u.ac.jp
†ishiwata.shintaro.es@osaka-u.ac.jp

**A strong coupling between electric polarization and elastic deformation in solids is an important factor in creating useful electromechanical nanodevices. Such coupling is typically allowed in insulating materials with inversion symmetry breaking as exemplified by the piezoelectric effect in ferroelectric materials. Therefore, materials with metallicity and centrosymmetry have tended to be out of scope in this perspective. Here, we report the observation of giant elastic deformation by the application of an alternating electric current in topological semimetals (V,Mo)Te$_2$, regardless of the centrosymmetry. Considering the crystal and band structures and the asymmetric measurement configurations in addition to the absence of the electromechanical effect in a trivial semimetal TiTe$_2$, the observed effect is discussed in terms of a Berry-phase-derived converse flexoelectric effect in metals. The observation of the flexoelectric effect in topological semimetals paves a way for a new type of nanoscale electromechanical sensors and energy harvesting.**

Electromechanical response in solids with broken inversion symmetry has been a central issue in condensed matter physics and extensively studied for industrial applications. A representative example is a piezoelectric effect observed in ferroelectric oxides[1]. Theoretical developments in recent decades have revealed that the electric polarization induced by these effects is dominated by the Berry phase of the electronic wavefunction [2–4]. Whereas such electromechanical responses have been observed mostly in insulators and some in semiconductors [5,6], metallic systems with the Berry phase can be promising candidates as well, when adopting alternating electric fields (currents) reducing the screening effect of charge carriers [7,8]. In fact, alternating-current-induced magnetopiezoelectric effects have been confirmed in metallic systems with magnetic ordering which breaks time and space inversion symmetries [9–14]. However, the piezoelectric response in metallic systems is typically small and has received less attention from the viewpoint of practical application.

Recently, the flexoelectric (FxE) effect, an electrical response to strain gradients breaking the inversion symmetry of the material, has attracted attention as a ubiquitous electromechanical response that can be observed in a wide range of materials with and without centrosymmetry [4,15–19]. In particular, topological materials with a finite Berry phase are expected to display a substantial FxE response [20,21]. In this perspective, transition-metal dichalcogenides studied as topological semimetals can be promising candidates showing a large electromechanical response including FxE effect [22,23]. They have a rich variety of polymorphic structures with and without inversion symmetry [24–27], thus showing various attractive quantum phenomena, typified by WTe$_2$, MoTe$_2$, and VTe$_2$ with Weyl or Dirac points [23,28–30].

Considering the lattice symmetry and the topological electronic structure, we focused on the three types of metallic transition-metal dichalcogenides 1T-TiTe$_2$, 1T''-VTe$_2$, and 1T'-MoTe$_2$, the latter two of which are topological semimetals. The crystal structures at room temperature are shown in Fig. 1a. Note that MoTe$_2$ undergoes a structural transition from the high-temperature centrosymmetric (1T') to the low-temperature polar (T$_d$) phase at about 250 K [25]. Whereas TiTe$_2$ is a trivial semimetal in terms of topological band structure [31], T$_d$-MoTe$_2$ and 1T''-VTe$_2$ are classified as the topological semimetals with Weyl and Dirac points, respectively [23,28–30]. Figure 1b shows the temperature dependence of in-plane resistivity ρ for (Ti,V,Mo)Te$_2$. ρ for all samples is metallic ($d\rho/dT > 0$) in the entire temperature regime, and the anomaly due to the polar-nonpolar structural transition is observed at 250 K for MoTe$_2$ [32–34]. In these materials, we demonstrate the current-induced mechanical response, characterized as the FxE effect, in metallic materials (Figs. 2a and 2b).

Figures 2c-e show the FFT spectra of the current-induced displacements of the sample surface for (Ti,V,Mo)Te$_2$ at 300 K with the application of the alternating current (AC current) of 100 mA, the maximum value that could be applied, in the selected frequency of 4 kHz (for details see Supplementary Figure 1). When the current is applied, a peaky signal is observed at 4 kHz for MoTe$_2$ and VTe$_2$, but not for TiTe$_2$. The typical magnitude of the displacement is 0.3 nm and 0.7 nm for MoTe$_2$ and VTe$_2$, respectively; the results of multiple measurements are shown in Supplementary Figures 2 and 3, where the MoTe$_2$ and VTe$_2$ exhibit the displacement values of 0.3-0.4 nm and 0.5-1.0 nm, respectively. These values are more than 10 times larger than the previously reported magneto-piezoelectric effects in metals[9–11], and are equivalent to semiconducting materials with larger electrical resistivity (~ 1 Ωcm)[35]. On the other hand, for TiTe$_2$ and MoTe$_2$, signals were observed at twice the frequency (8 kHz) of the applied current, which can be assigned as Joule heating as confirmed below (the slight Joule heating effect is observed also in VTe$_2$ as shown in Supplementary Figure 3). The observed sample dependence of the Joule heating effect can be attributed to differences in the way of energy consumption: in VTe$_2$, most of the input energy is converted to strain response, while in TiTe$_2$, most of the input energy is consumed not as strain effect but as the Joule heating. It is noteworthy that the gigantic displacement signals are detected in MoTe$_2$ and VTe$_2$ with a topological band structure.

To get more insight into the electromechanical response, we measured the temperature dependence of displacement signals from 80 to 300 K for (Ti,V,Mo)Te$_2$ as shown in Figs. 3a and 3b. The displacement signals at 4 kHz corresponding to the electromechanical response increase with increasing the temperature for MoTe$_2$ and VTe$_2$. In contrast, the Joule heating signals at 8 kHz are nearly independent from temperature. It is noted here that the electromechanical signals are observed not only in the low-temperature polar phase but in the high-temperature nonpolar phase of MoTe$_2$, indicating that the electromechanical signals at 4 kHz are observable regardless of the inversion symmetry of the crystal. This result suggests that the present signal has a distinct origin from the previously reported piezoelectric effects in noncentrosymmetric materials[8,35]. The temperature dependences for VTe$_2$ and MoTe$_2$ are likely related to the change in the resistivity as reported for semiconducting materials[35], which will be discussed later. We further investigate the properties of the current-induced displacement signals at 300 K. Figures 3c and 3d show the current dependence of the displacement signals using the AC current of 4 kHz. The magnitude of the signals at 4 kHz in MoTe$_2$ and VTe$_2$ increases with increasing current amplitudes, and notably, the increase rate is linearly proportional to the amplitude of the electric current. In contrast, the signals at 8 kHz in TiTe$_2$ and MoTe$_2$ are proportional to the square of the current magnitude $I^2$, characteristic to a relative change in the length of the sample $\Delta L/L$, which can be related to the Joule heating ($\propto I^2$) as $\Delta L/L = \alpha\Delta T \propto C\Delta T \propto I^2$. Here the linear expansion coefficient $\alpha$ and the heat capacity $C$ are assumed to be almost temperature-independent, especially around 300 K. These current dependencies support that the signal at 4 kHz is an intrinsic electromechanical response rather than a heating response.

Here, we discuss the origin of the current-induced electromechanical response in metallic systems. For the metallic systems, it is proposed that breaking the space-inversion symmetry leads to dynamical distortion in response to AC currents[8]. In particular, the dynamical response for magnetic systems due to the magnetic-order-induced simultaneous breaking of the time-reversal and space-inversion symmetry (magnetopiezoelectric effect) has been reported[9–13]. However, the current-induced electromechanical response observed in (V,Mo)Te$_2$ requires neither magnetic order nor broken inversion symmetry. Therefore, the possibility of the dynamical (magneto)piezoelectric effect can be excluded as the origin of the electromechanical response in (V, Mo)Te$_2$. Another possible origin for the observed electromechanical response in (V, Mo)Te$_2$ is the converse FxE effect, the opposite effect of the FxE effect. The converse FxE effect manifests itself as the asymmetric crystal distortion induced by the application of an asymmetric electric field. In this case, an electromechanical response can be observed, irrespective of the centrosymmetry of the crystal, when the applied electric field or current breaks inversion symmetry. In our measurement, the top and bottom electrodes of the crystal are attached asymmetrically, as shown in Fig. 2a, which causes a spatial variation in the current flowing, potentially resulting in the observation of the converse FxE effect in (V, Mo)Te$_2$. This inhomogeneous current flow is also supported by the position dependence of the displacement signal (distance from the electrode). Figures 4(a) and (b) exhibit the displacement signals measured at two different positions on the crystals of VTe$_2$ and MoTe$_2$, respectively. In this measurement, a large signal is observed at a position close to the electrode, while the signal disappears immediately when the measurement position is away from the electrode (the results of the measurements with different samples and higher frequency resolution are shown in Supplementary Figure 4). To our knowledge, the converse FxE effect, which has been reported for insulating and semiconducting systems, in this study is the first observation in a metallic system.

The synchronized displacement measured at the fixed current of 100 mA is shown as a function of the frequency of the applied AC current for all samples in Fig. 4c. These strong frequency dependencies in the kHz range are observed in the semiconducting material[35], but in stark contrast to the cases for conventional (insulating) piezoelectric materials[36]. The inset of Fig. 4c exhibits the log-log plot of the displacement signals for MoTe$_2$ and VTe$_2$. In the low-frequency regime below 0.3 kHz, the displacement signal rapidly decreases, possibly reflecting the polarization screening by conduction electrons. On the other hand, in the high-frequency regime, roughly the $f^{-1}$ dependent signal is observed, which is characteristic to the converse FxE effect in metallic systems as described below. Considering the phenomenological equations for the converse FxE effects[37], we describe the relationship between the electric current and the electromechanical signal. The converse FxE effect can be observed in centrosymmetric crystals when the mirror symmetry is broken by a nonuniform field (current). The relation of the polarization $P$ to the nonuniformity of the strain field $S$ (stress $X$) and that of the strain (stress) to the nonuniformity of the electric field $E$, are known as FxE and converse FxE effect, respectively: $P_l = \mu_{ijkl}(dS_{ij}/dx_k)$ [$P_l = F_{ijkl}(dX_{ij}/dx_k)$] and $X_{ij} = -\mu_{ijkl}(dE_l/dx_k)$ [$S_{ij} = F_{ijkl}(dE_l/dx_k)$], where $\mu_{ijkl}$ and $F_{ijkl}$ are the components of the so-called (converse) FxE tensors. While the piezoelectric property is nonzero only for noncentrosymmetric materials, the (converse) FxE tensors are, in principle, nonzero for all materials.

In conducting materials under the AC current (field), there are two types of electrical current density components $j_e$ and $j_D$, where $j_e$ and $j_D$ stem from the conduction electrons and the time variation of electric flux density $D$ associated with the electric polarization ($D=\varepsilon_0 E+P$), respectively. The $j_D$ component can be present even in the low-frequency region, where the $j_e$ component is generally dominant. $j_D$ is described by the following equations: $j_D = dD/dt = d(\varepsilon_0 E+P)/dt$. Given that the applied AC current (current density) is expressed as a sinusoidal form as $j_{ac}(t) = j_0\sin\omega t$, the strain is roughly described by

$S_{ij}(t) = F_{ijkl}/\sqrt{[(1/\rho)^2+(\omega\varepsilon)^2]}\sin\omega t \sim \omega^{-1}\sin\omega t$  (1),

which is derived from the converse FxE equation of $S_{ij} = F_{ijkl}(dE_l/dx_k)$. Here we assume that the time $t$ and position $x$ are independent variables (as discussed in Supplementary Note 5). The equation (1) indicates that the displacement signals of the converse FxE response $S(t)$ in conducting materials should be inversely proportional to the frequency $f$ as well as the angular frequency $\omega$ under the AC current as shown in the inset of Fig. 4c.

Another characteristic of the present model of the FxE effect is that it is roughly proportional to electrical resistivity $\rho$ [$S_{ij}(t) \sim F_{ijkl}/\sqrt{[(1/\rho)^2+(\omega\varepsilon)^2]}\sin\omega t \sim \rho\sin\omega t$]. In fact, both the electrical resistivity and displacement signal for VTe$_2$ and MoTe$_2$ decrease monotonically with decreasing temperature. Furthermore, in MoTe$_2$, a thermal hysteresis associated with the structural phase transition is observed in the displacement signal as well as in the electrical resistivity (see Supplementary Figure 7).

Although the FxE effect can, in principle[7], be observed in all materials, it was absent in TiTe$_2$ while it was observed as a gigantic effect in MoTe$_2$ and VTe$_2$. The absence of FxE effect in TiTe$_2$ can be associated with its fairly low resistivity as compared with the semiconducting material such as AgCrSe$_2$[35]. However, it is notable that MoTe$_2$ and VTe$_2$ show a gigantic FxE effect in spite of the low electrical resistivity comparable to that of TiTe$_2$. Theoretically, an electric current response caused by spatially asymmetric strain, such as the FxE effect, is predicted in topological semimetals[20,21]. This FxE-like effect depends on the magnitude of the second-class Chern-flux, which is derived from the second Chern-form of the Berry curvature. This component of the Berry curvature stems from the momentum *k* and the spatial position *r* dependencies of the Bloch wave function and energy dispersion in the presence of spatial inhomogeneity. Therefore, as a converse FxE effect, large electromechanical responses with respect to asymmetric electric currents can be expected in topological semimetals such as the Weyl semimetal $T_d$ phase in $MoTe_2$ at low temperatures and the Dirac semimetal $VTe_2$, not in the trivial semimetal $TiTe_2$. However, we should note that large FxE responses are also observed in the high-temperature phase of $MoTe_2$, of which band structure is expected to be topologically trivial. As an origin of the observation of the large converse FxE effect in the trivial phase of $MoTe_2$, we consider the effect of interband hopping of electrons, which is allowed as a consequence of the temperature broadening of the Fermi distribution function. This mechanism is similar to the shift current photovoltaic effect in ferroelectric materials, where the Berry phase contribution arises from the electron excitation between the conduction and valence bands.[38,39] Provided that this is the case for the trivial phase of $MoTe_2$, the FxE responses due to polarization currents can emerge even in a metallic state, which potentially leads to pioneering electromechanical functions in nearly topological materials. On the other hand, $MoTe_2$ and $VTe_2$ have structural instabilities associated with a nonpolar-polar structural transition and a CDW transition, respectively, which can be another factor amplifying the FxE response. In addition, whereas a current-induced electromechanical response, such as the dynamical piezoelectric effect and current-induced strain effect is expected for multipole order involving polar symmetry breaking, it is likely that such effects tend to be negligibly small due to the presence of multi-domains[8,40]. To further characterize the observed electromechanical effect, it is indispensable to perform the measurements of local current density and local displacement signal in addition to the development of a microscopic theory of the converse FxE effect in metallic systems.

In conclusion, we conducted the current-induced displacement measurements for the three types of transition-metal dichalcogenides $(Ti,V,Mo)Te_2$ with different structural and electronic features. Large current-induced lattice displacements synchronized with the frequency of the applied AC current, which was assigned as intrinsic electromechanical effect, were observed in the topological semimetals $MoTe_2$ and $VTe_2$, but not in a trivial semimetal $TiTe_2$. On the other hand, the lattice displacement derived from Joule heating was negligibly small in $MoTe_2$ and $VTe_2$, while it was significant in $TiTe_2$. Considering the facts that the displacement follows $f^{-1}$ dependence in the high-frequency regime and depends not on the inversion symmetry of the crystal but on the measured position on the crystal, the observed electromechanical effect can be characterized as the converse FxE effect in metallic systems. The observation of FxE-like response in topological semimetals in this study would expand a strategy for the development of new types of electromechanical materials for micro sensors and power generation devices.

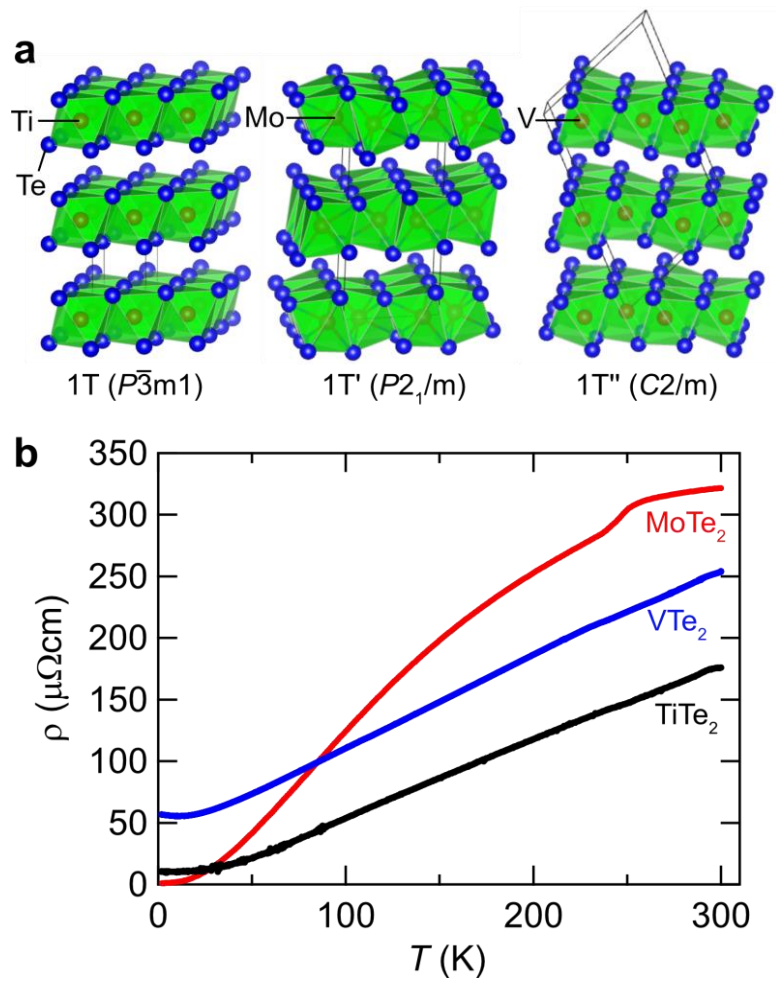

**Fig. 1 Crystal structures and electrical resistivity. a,** Crystal structures of transition-metal dichalcogenides [1T-TiTe$_2$, 1T'-MoTe$_2$, and 1T''-VTe$_2$] at room temperature. **b,** Temperature dependence of electrical resistivity ρ.

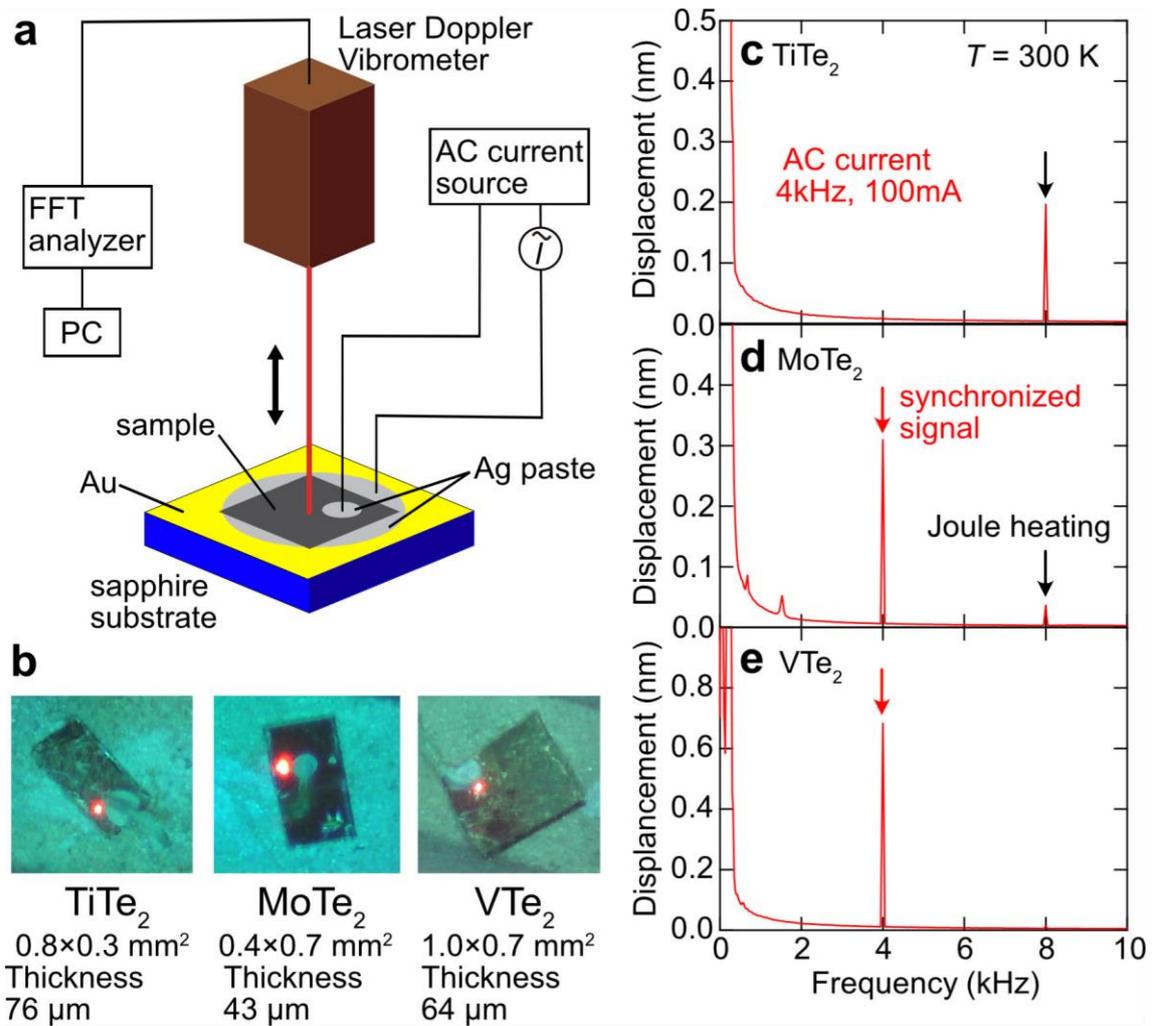

**Fig. 2 Measurement setup and displacement signals. a,** Schematic illustration of measurement setup for the dynamical flexoelectric (piezoelectric) response. **b,** Photographs of the single crystals for the measurements. Frequency-dependent spectra of displacement signals by application of AC current of 100 mA with a frequency of 4 kHz at 300 K for (**c**) TiTe$_2$ (**d**) MoTe$_2$, and (**e**) VTe$_2$. AC displacement signals appear at the same frequency as the AC electric current as denoted by red arrows in the figures. Thermal expansion signals caused by Joule heating are observed at twice the frequency of AC current as denoted by black arrows.

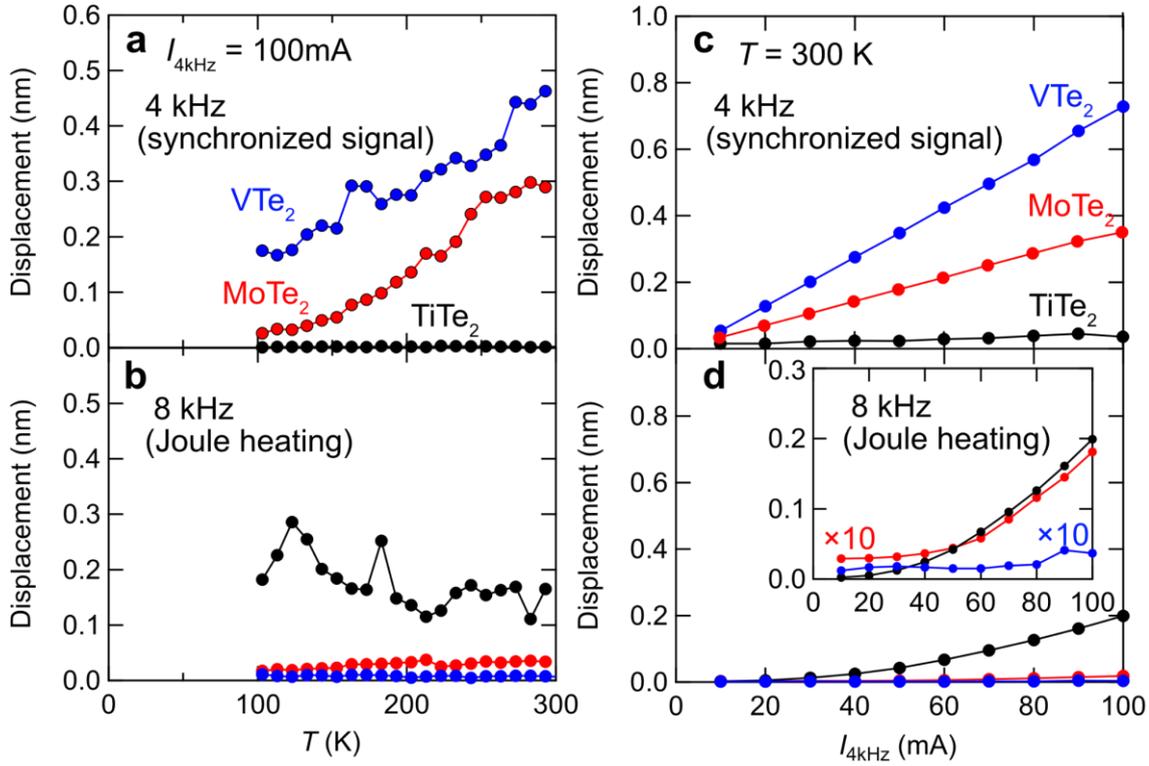

**Fig. 3 Temperature and current dependence of displacement signals.** Temperature dependence of (**a**) the displacement signal measured at 4 kHz and (**b**) the Joule heating signal at 8 kHz using AC current with 100 mA and 4 kHz for TiTe$_2$, MoTe$_2$, and VTe$_2$. Current dependence of (**c**) the displacement signal measured at 4 kHz and (**d**) the Joule heating signal at 8 kHz using AC current with 4 kHz at 300 K. Inset shows the magnified figure for the Joule heating signal.

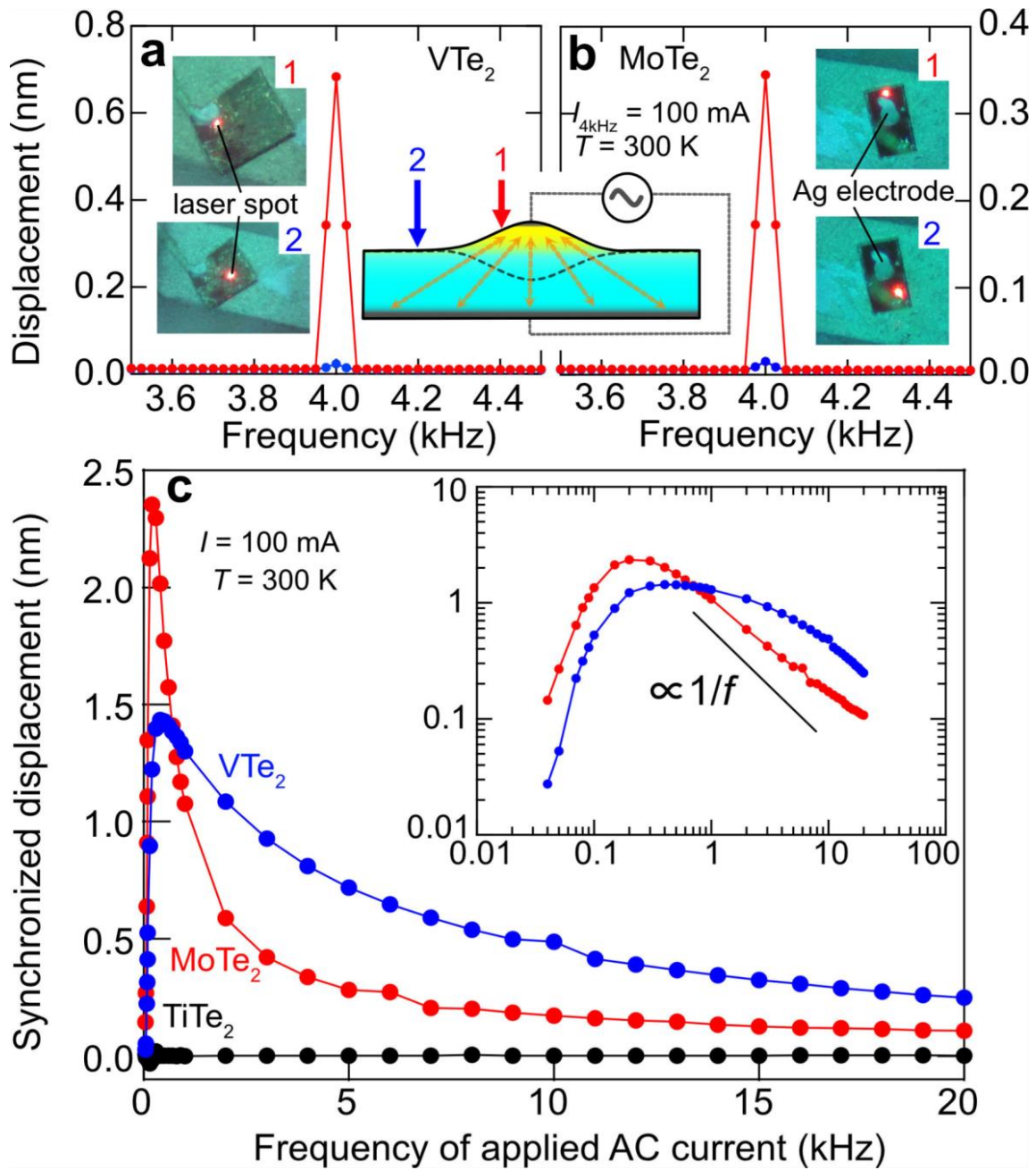

**Fig. 4 Position and synchronized frequency dependence of the displacement signals.** The displacement signals for (a) VTe$_2$ and (b) MoTe$_2$ at different positions of the laser spot, 1 and 2. The displacement signals become smaller as the laser spot position moves away from the electrode. The inset pictures show the positions of the laser spot and the Ag electrode. The inset illustration schematically represents the displacement of crystals induced by the applied AC current. (c) Synchronized displacement signal for TiTe$_2$, MoTe$_2$, and VTe$_2$ as a function of the frequency of the applied AC current with 100 mA at 300 K. These data are prepared from the frequency dependence of the displacement signals measured with selected frequency of applied AC current varied from 0 to 20 kHz (see Supplementary Figure 5). Inset shows the log-log plot for MoTe$_2$ and TiTe$_2$.

## Methods

### Single crystals of (Ti,V,Mo)Te$_2$

Single crystals of (Ti,V,Mo)Te$_2$ were grown using the NaCl-flux method[34]. The single crystals were confirmed to be a single phase by a RigakuXtaLAB-mini II diffractometer with graphite monochromated Mo $K\alpha$ radiation. In addition, the structural transition of VTe$_2$ from 1T to 1T'' between 436 K and 500 K was identified by the synchrotron X-ray diffraction measurements for single crystalline samples at BL02B1 in SPring-8 as shown in Supplementary Figure 8.

### Measurement for displacement signals

The Plate-like single crystals were fixed to a gold-deposited sapphire substrate using the Ag paste. The sample sizes are shown in Fig 2b. Current electrodes were formed on the samples, as shown in Fig. 2a. While applying the AC current to the samples along the $c$ directions, time-dependent displacements generated along the $c$ direction were measured using a laser Doppler vibrometer combined with FFT analyzer (Ono Sokki). A red laser is directed at the surface of (Ti,V,Mo)Te$_2$, and the vibration velocity of the sample is extracted from the Doppler shift of the reflected laser. The observed velocity was then numerically integrated with respect to time using the FFT analyzer to obtain the vibration amplitude of the sample. An objective lens's laser spot diameter is less than 100 microns. The temperature dependence was measured by the temperature control system THMS600 (Linkam Scientific Instruments). The sample size and pictures in the measurements are shown in Fig. 2b. Since the magnitude of the displacement signal strongly depends on the measurement position, the absolute value changes due to slight deviations in the respective measurement position for the temperature dependence, frequency dependence, and current dependence. During the measurement of each dependence, positional deviations are almost negligible.

**Acknowledgements**
The authors thank H. Matsuura and M. Ogata for fruitful discussions. This study was supported in part by KAKENHI (Grant No. JP20K03802, JP21H01030, JP21K13878, JP22H00343, JP23H04871 and JP24K00570), FOREST (No. JPMJFR236K) from JST, Murata Foundation, Yazaki Memorial Foundation for Science, Technology, and Asahi Glass Foundation.